# Boosting the Memory Window of Memristive Stacks via Engineered Interfaces with High Ionic Mobility

Noa Varela-Domínguez, Daniel Veira-Canle, Nicholas Davey-García, Victor Leborán, Rafael Ramos, Fèlix Casanova, Luis E. Hueso, Victor M. Brea, Paula López, Francisco Rivadulla, and José Diogo Costa*

**ABSTRACT:** Realizing the potential of oxide-based memristive devices for high-density data storage and energy-efficient computing still relies on overcoming key technical challenges, including the need for a larger number of stable resistance states, faster switching speeds, lower SET/RESET voltages, improved endurance, and reduced variability. Addressing these limitations requires innovative material design strategies. Here, we demonstrate that introducing a thin layer of oxide-ion conductor $SrCoO_{3-x}$ between the metal and the $SrTiO_3$-based memristive elements, expands the number of distinguishable resistance states from ≈8 to ≈22. This modification also reduced the SET/RESET voltage by 50% and markedly improved device endurance, albeit with a trade-off of reduced state retention. To assess the performance of this architecture, we trained a two-layer fully connected neural network using the experimental $SrTiO_3/SrCoO_{3-x}$ memristor characteristics on the MNIST handwritten digit dataset. Networks with hidden-layer sizes between 64 and 256 neurons achieved classification errors below 7%. Finally, we confirmed the transferability of this interface-engineering approach by applying it to $HfO_x$-based devices, achieving a consistent enhancement in the resistive state window.

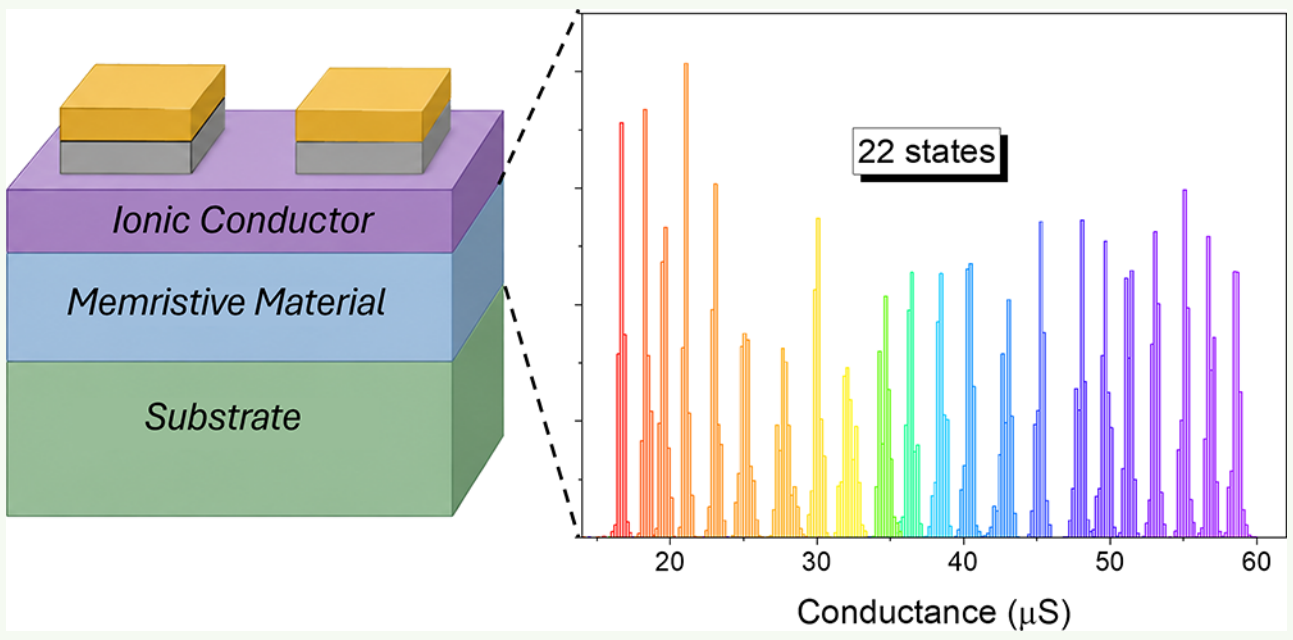

## I. INTRODUCTION

The rapid growth of data-driven technologies and artificial intelligence has placed unprecedented demands on modern computing architectures. These are increasingly constrained by the separation of memory and logic units, leading to bottlenecks in speed and energy efficiency. As a result, there is a pressing need for novel hardware paradigms that can support high-density, low-power, and scalable data processing. Among emerging technologies, resistive switching (RS) devices have gained substantial attention for integrating memory and computational units within a single, compact structure.[1−4]

RS devices operate by toggling between distinct resistance states under applied electrical stimuli, enabling them to store and process information in a tendentially non-volatile manner. While binary switching has been the conventional focus,[5,6] recent interest has shifted toward multi-level resistive switching, wherein a single device can access multiple discrete conductance states.[7−9] This capability is especially valuable for multilevel memory storage, analog computing, and neuromorphic systems, since fine control over conductance enables increased bit density. More precisely, while a binary (2-state) system can encode only a single bit (0 or 1), a 4-state system can encode 2 bits (00, 01, 10, 11), with the general pattern following a $2^N$ relationship. Achieving stable and tunable multistate behavior, however, remains a significant materials challenge, often limited by instabilities at the electrode/dielectric interface, stochastic switching dynamics, or poor device endurance, particularly in valence-change resistive switching oxides.

Important advances in this regard have focused on exploring dielectric systems that accommodate high oxygen deficiency without phase breakdown—such as $TiO_x$, $HfO_x$, $HfAlO_x$, $TaO_x$, or $Sr(Ti,Fe,Co)O_3$,[7,10−14] or the use of nanocomposites.[15] Complementary strategies involve interface engineering to modulate oxygen exchange, for instance, by introducing ion-migration-blocking layers like as SrO or $Al_2O_3$.[16−18] While these layers effectively confine ionic transport and enhance retention, they often impose significant trade-offs: the additional barrier increases operating voltages and power consumption, while the increased series resistance can induce local Joule

**Received:** June 1, 2026
**Revised:** August 20, 2026
**Accepted:** August 21, 2026

heating, accelerating dielectric breakdown and degrading endurance.

Here we propose an alternative strategy, in which an oxygen-conductive interfacial layer of $SrCoO_{3-x}$ (SCO) is inserted between the oxide and the metal electrode to facilitate the oxygen exchange and work as an oxygen reservoir. We show that this strategy expands the range of distinguishable resistance levels from 8 states (3-bit equivalent) in the $SrTiO_3$ (STO) reference device, to 22 states in the STO/SCO heterostructure, surpassing the threshold for 4-bit encoding. Equivalently, this translates into an expanded conductance range, thereby enhancing the operational range of synaptic[19,20] or reservoir computing systems.[21,22]

We demonstrate the generality of this strategy by extending it to a CMOS-compatible $HfO_x$ platform. The incorporation of the SCO interfacial layer produces a comparable expansion in the resistive state window and number of distinguishable resistance states, showing the robustness of the approach and its suitability for scalable processes.

## II. EXPERIMENTAL/MATERIALS AND METHODS

Two sets of thin films of $SrTiO_3$ (STO; 25.5 nm) were deposited by pulsed laser deposition (PLD) on (001)-oriented Nb-doped (0.5 wt %) $SrTiO_3$ (Nb:STO) substrates, at 1.3 J/cm$^2$ and 765 °C under an oxygen partial pressure of 100 mTorr. An additional layer of $SrCoO_{3-x}$ (SCO; 5 nm; deposited at 750 °C under the same laser fluence and oxygen pressure) was grown on top of STO in some samples.

After deposition, top electrodes (Pt 5 nm / Au 60 nm) were patterned using standard photolithography and lift-off. The whole process leads to samples of composition (1) Nb:STO/STO/Pt/Au (STO devices from now on) and (2) Nb:STO/STO/SCO/Pt/Au (STO/SCO devices from now on). The structural properties of the STO and STO/SCO were characterized using X-ray diffraction and transmission electron microscopy (TEM).

A second series of devices was fabricated to evaluate the transferability of this strategy, consisting of (1) Nb:STO/$HfO_x$ (25.5 nm) and (2) Nb:STO/$HfO_x$ (25.5 nm) / SCO (5 nm) heterostructures. These films were deposited at 100 mTorr and 400 °C for both the $HfO_x$ and SCO layers.

Electrical characterization was carried out at room temperature under ambient atmospheric conditions using a Keithley 2450 source meter in a two-terminal configuration, with the Nb:STO substrate acting as the bottom electrode. Fast pulsed measurements down to 5 μs were performed using a TGA 1240 waveform generator. All data shown correspond to devices with 55 μm × 55 μm pads unless otherwise specified.

## III. RESULTS AND DISCUSSION

### A. Structural and Chemical Characterization

Figure 1(a) summarizes the microstructural characterization of the STO/SCO devices. The high-resolution TEM imaging and Energy Dispersive X-ray Spectroscopy (EDS) analysis confirm a well-defined compositional separation between the STO and SCO layers. Both films are crystalline, with a smooth interface and good epitaxial matching to the substrate.

Furthermore, the XRD θ–2θ scans of both STO and STO/SCO samples [Figure 1 (c)] display a distinct shoulder close to the sharp (002) Nb:STO substrate reflection, as well as well-defined thickness fringes. This further confirms close to stoichiometric growth of STO, and smooth interfaces. Note that due to the very thin nature of the SCO layer and the epitaxial strain, the precise oxygen stoichiometry cannot be unequivocally determined from these measurements.

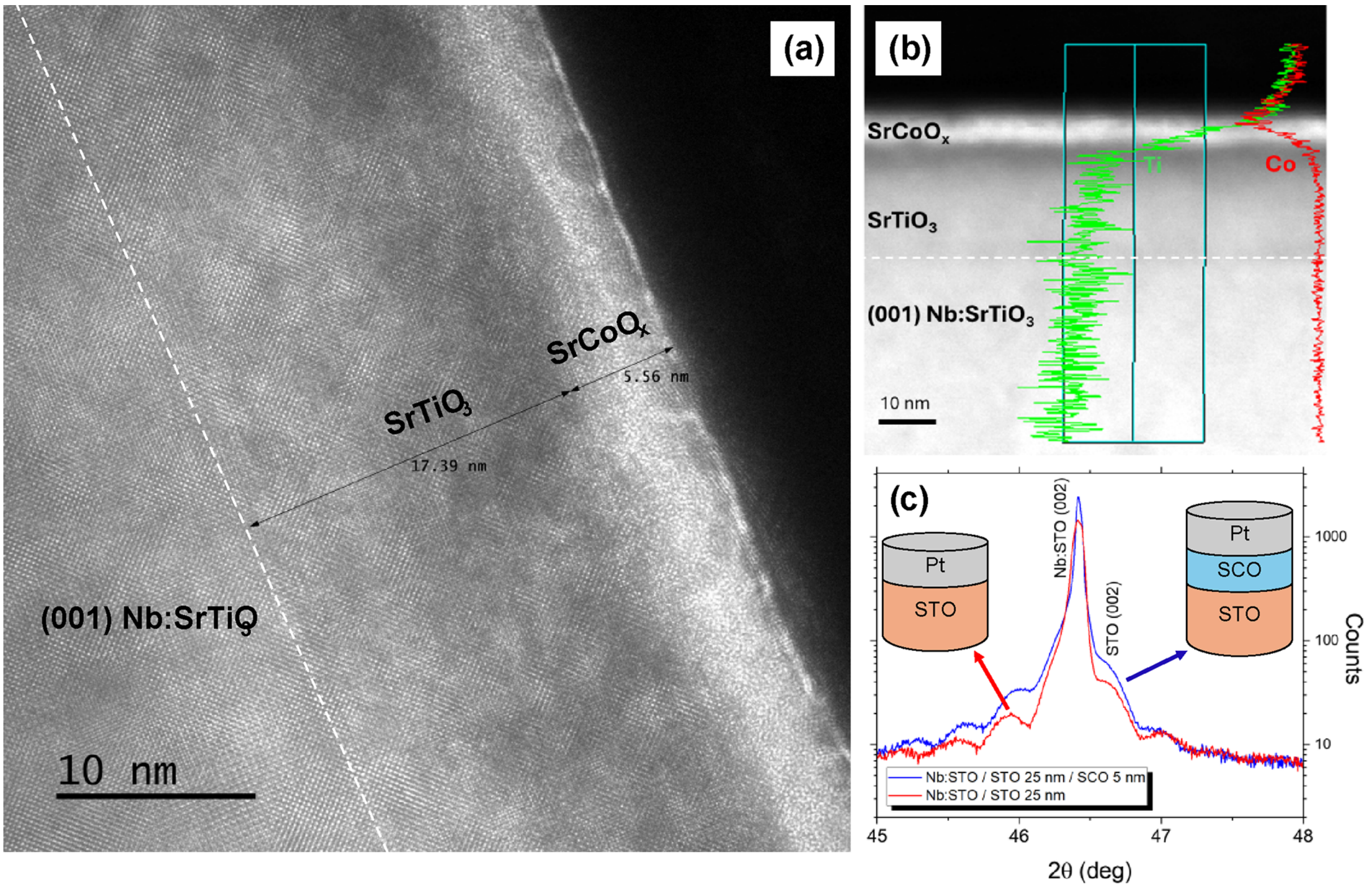

**Figure 1.** (a) High-resolution TEM imaging and (b) EDS analysis of a cross-sectional lamella of a STO/SCO device. (c) High-resolution XRD structural characterization of the STO (red line) and STO/SCO (blue line). The inset shows the deposited stack, including the Pt electrical contact.

## B. DC *I*–*V* Characterization

The RS behavior of STO and STO/SCO devices is shown in Figure 2(a). Both show eightwise switching[23] (black arrows indicate the sweeping direction), due to the accumulation/depletion of oxygen vacancies ($V_{\ddot{O}}$) near the dielectric/electrode interface under positive/negative voltage. Nevertheless, both devices also show notable differences: (i) the forming voltage of STO/SCO (~3.6 V), is substantially lower than for STO (~4.8 V), suggesting an easier filament formation due to the enhanced ionic mobility interface (Figure S1, Supporting Information); (ii) a larger electrical conductivity of STO/SCO, manifested as a larger current at the same voltage; (iii) more importantly, significant and irreversible damage was observed in the STO sample after only a few cycles at 3 V (see also Figure S4, Supporting Information), whereas the STO/SCO devices showed no measurable degradation under the same cycling conditions. Similarly, endurance tests over $10^5$ cycles using ±2 V pulses reveal pronounced degradation in the STO sample, whereas the STO/SCO sample maintains stable performance throughout this extensive cycling procedure [Figure 2](b).

This enhanced stability translates into markedly improved reproducibility, characterized by lower device-to-device variability and extended cyclability without performance loss (Figures S3 and S4). From now on, and for the sake of comparisons between both types of devices, we limited the operating voltage to 2 V.

## C. RF Performance Analysis

Radiofrequency (RF) electrical pulses are commonly used to characterize memristors, as they replicate the real operating conditions encountered in practical applications. By applying RF pulses, different resistance states can be induced in the device, allowing for a more accurate assessment of its switching behavior, retention capabilities, and endurance. The devices were programmed using 50 ms SET/RESET pulses at ±2 V, and the resulting resistance states were measured with 50 ms read pulses at 0.5 V (see Supporting Information, Figure S5).

Figure 3(a,b) present the resistance states measured for the STO and STO/SCO devices, respectively. A pronounced enhancement in both the memory window and the number of discernible resistance states is observed, increasing from 8 in STO to 22 in STO/SCO devices. This remarkable improvement is consistent with the broader range of the *I*–*V* characteristics for the STO/SCO device, as shown in Figure 2 (note the logarithmic scale). Although direct comparison between different multilevel memristors should be approached with caution, since the reported number of states strongly depends on the programming protocol, read conditions, acceptance criteria, and retention requirements, we have added a comparison table in the Supporting Information (Table S1) summarizing representative literature reports.

We also extracted the SET/RESET voltages of the fabricated devices by performing a full SET/RESET cycle followed by a read–write–read sequence (see Supporting Information, Figure S6). Figure 4(a) shows a substantial reduction, close to 50%, in both SET and RESET voltages for the STO/SCO sample, with pulse widths down to 5 μs. This reduction indicates that the enhanced ionic mobility at the SCO interface facilitates lower operating voltages and enables shorter pulse operation, thereby reducing power consumption.

Additionally, the STO/SCO sample exhibits a lower RESET voltage than SET voltage, while the STO-only sample displays the opposite behavior. This is likely due to the facilitated ionic drift driven by the interfacial SCO layer, resulting in a stronger tendency for the conductance to relax toward the high resistance state (HRS), as discussed later. Since this spontaneous ionic drift occurs in the same direction as the RESET process, the transition from the low resistance state (LRS) to the HRS is facilitated, resulting in a slightly higher SET voltage (compared to RESET voltage). In contrast, the STO-only device exhibits a less pronounced ionic drift, resulting in a slightly higher RESET voltage. This asymmetry further supports the significant interfacial role of the SCO layer in enhancing oxygen exchange.

The microscopic picture of these systems can be understood by taking into consideration that $Ti^{4+}$ is in a $3d^0$ state in the stoichiometric STO layer, which is a band insulator with a bandgap of 3.2 eV, but can be readily transformed into an itinerant electron system with high mobility via $Ti^{4+} \rightarrow Ti^{3+}$ reduction.[24] This transition can be induced either by $Nb^{5+}$ doping or the introduction of oxygen vacancies. In contrast, SCO is a negative charge-transfer material, meaning that the band derived from the metal-oxygen antibonding interaction may cross the non-bonding O 2p states as the oxidation state of Co approaches 4+.[25,26] Therefore, the interface between STO and SCO must be electrochemically active, with spontaneous electron transfer from Ti to Co.

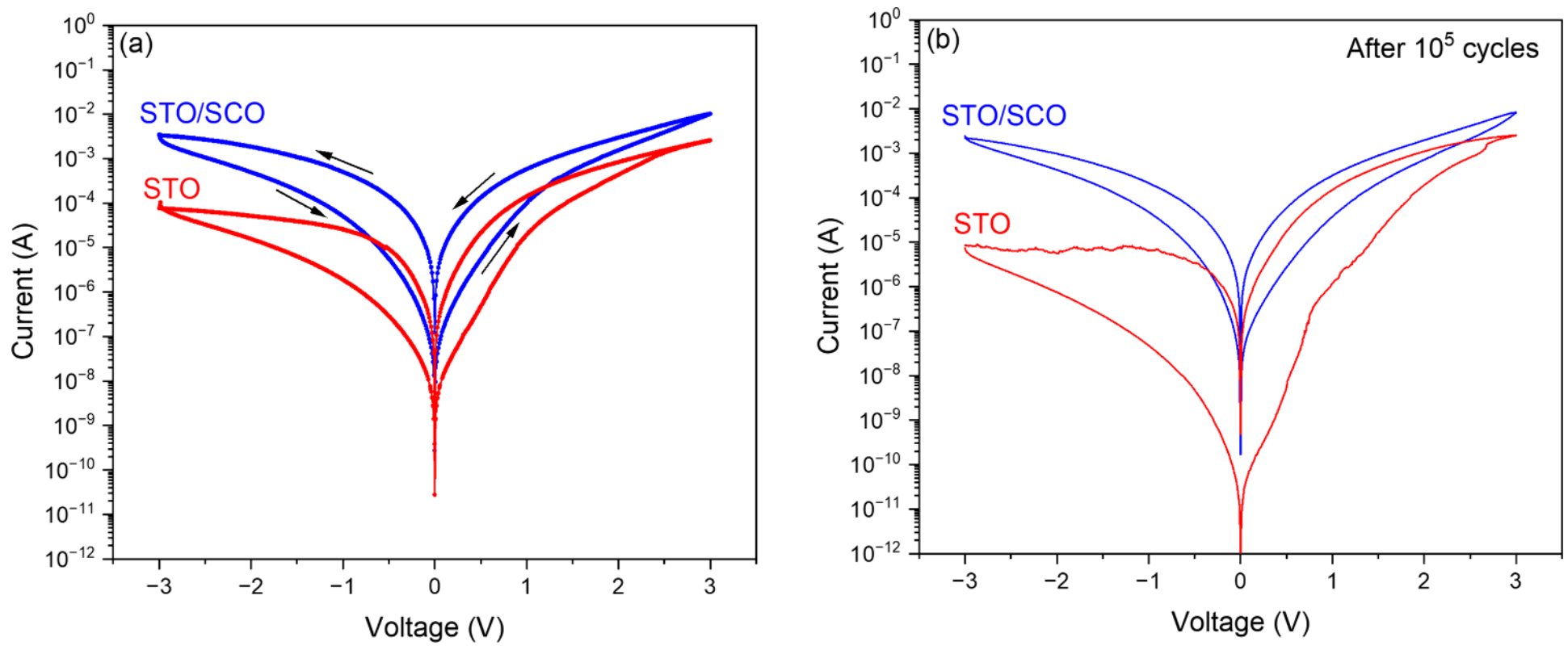

**Figure 2.** Comparison of the DC electrical performance of the devices of STO/SCO and STO. (a) Apart from the higher conductivity of the STO/SCO, with black arrows indicating the sweep direction; (b) the devices suffer almost no variation after $10^5$ cycles, while strong degradation is observed in STO devices.

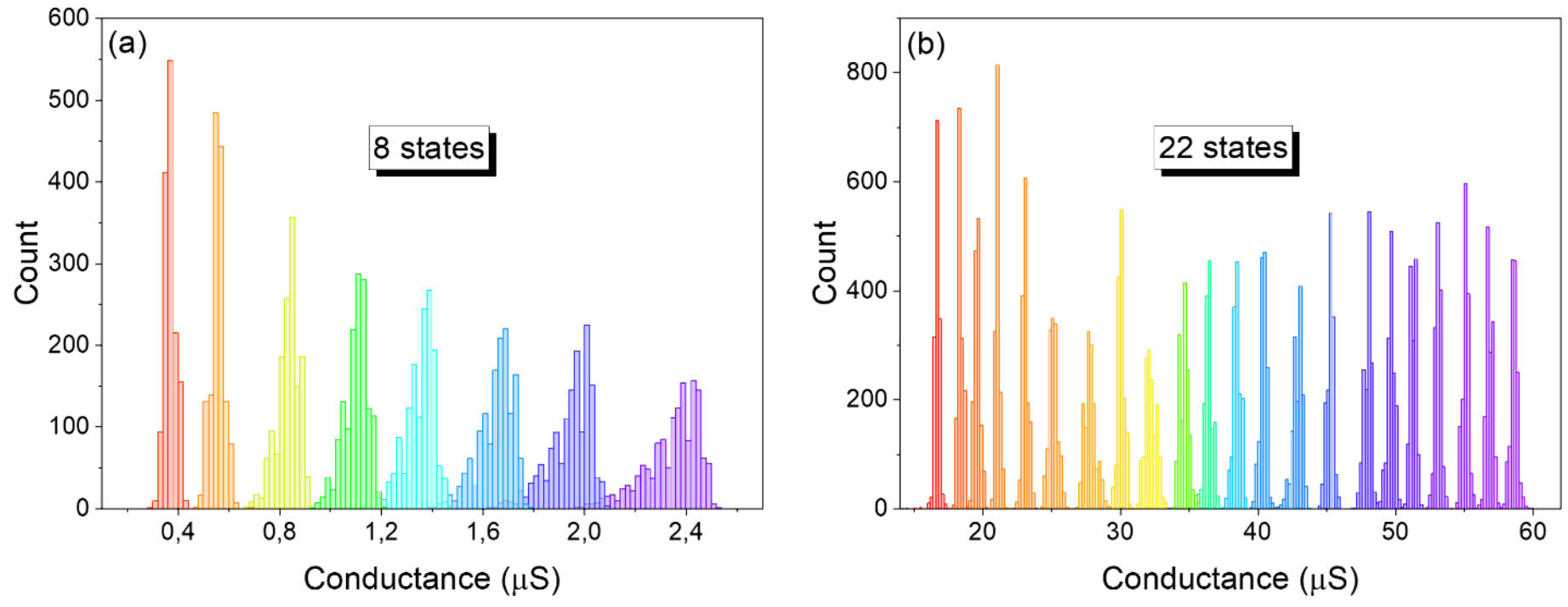

**Figure 3.** Current histograms showing distinguishable resistance states after a sequence of 50 ms-wide SET/RESET pulses, from highest to lowest achievable resistance states, for the (a) STO and (b) STO/SCO samples.

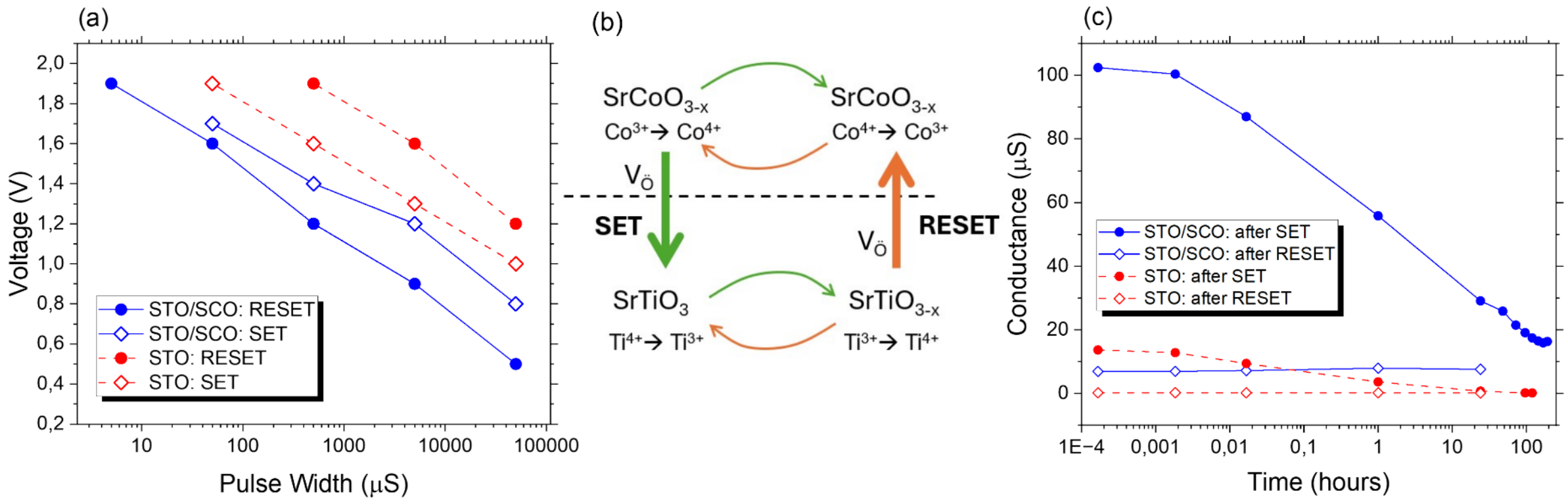

**Figure 4.** (a) SET (open diamonds) and RESET (filled circles) voltages as a function of pulse width for STO/SCO (solid blue line) and STO (dashed red line). (b) Schematic representation of the microscopic picture at the STO/SCO interface. (c) Measured conductance at the read voltage in the LRS (open diamonds) and HRS (filled circles) as a function of time for STO/SCO (solid blue line) and STO (dashed red line).

Thus, when a positive voltage is applied to this interface during the forming process, oxygen vacancies ($V_{\ddot{O}}$) are forced across the interface from SCO to STO, leading to a partial oxidation/reduction of SCO/STO, respectively. Since the perovskite phase of SCO (oxygen-rich) is more conductive than the brownmillerite phase (oxygen-deficient), and oxygen-deficient STO is more conductive than stoichiometric STO, the conductivity of both layers increases in comparison to the STO sample, consistent with the higher current observed in Figure 2. When the voltage is reversed, $V_{\ddot{O}}$ are driven back towards the SCO side of the interface, rendering it more insulating and increasing the resistance of both layers. The process of oxygen exchange across the SCO/STO interface is summarized in Figure 4(b).

Importantly, when the system is allowed to relax from the LRS at zero voltage, the HRS should be recovered due to the favorable oxidation of $Ti^{3+}$ in the presence of $Co^{4+}$ at the STO/SCO interface. To demonstrate this hypothesis, we measured the temporal evolution of the conductance in the STO/SCO sample following a full SET and a full RESET, spanning from seconds to multiple days [Figure 4(c)]. As expected, a pronounced drift toward the HRS is observed over time, consistent with spontaneous oxygen exchange at the SCO/STO interface. Nevertheless, the application of a subsequent full SET pulse restores the device to a conductance state close to its initial value. The STO-only device also exhibits a gradual drift toward the HRS, eventually approaching pre-forming resistance values on a timescale of days. However, unlike the STO/SCO device, applying a full SET pulse at this stage does not recover the original state, and a new forming cycle is required. It is worth noting that although long retention times are generally desirable for conventional data-storage applications, short-term memory characteristics can be advantageous for other computing paradigms, such as reservoir computing.[26]

## D. Transferability to Other Material Frameworks

The proposed approach is promising as a material heterostructure for increasing the number of resistance states in STO-based memristors. To assess the generality of this approach, we selected $HfO_x$, which is the most widely used memristive material, as an alternative valence-change-based system. Following the same methodology as in the previous sections, two sets of samples were fabricated: (i) $HfO_x$ (25.5 nm) and (ii) $HfO_x$ (25.5 nm)/SCO (5 nm), as described in the experimental section.

The comparison of the *I*–*V* curves of both devices in Figure 5(a) shows a current enhancement in $HfO_x$/SCO with respect to STO, similar as STO/SCO discussed in Figure 2. It should be noted that there was a more significant dependence on the pad size in the $HfO_x$ sample, and the results presented here concern 220 μm pad size. This is an indication of a different mechanism for filament formation that, nevertheless, does not affect the improvement in performance achieved by the introduction of the SCO layer.

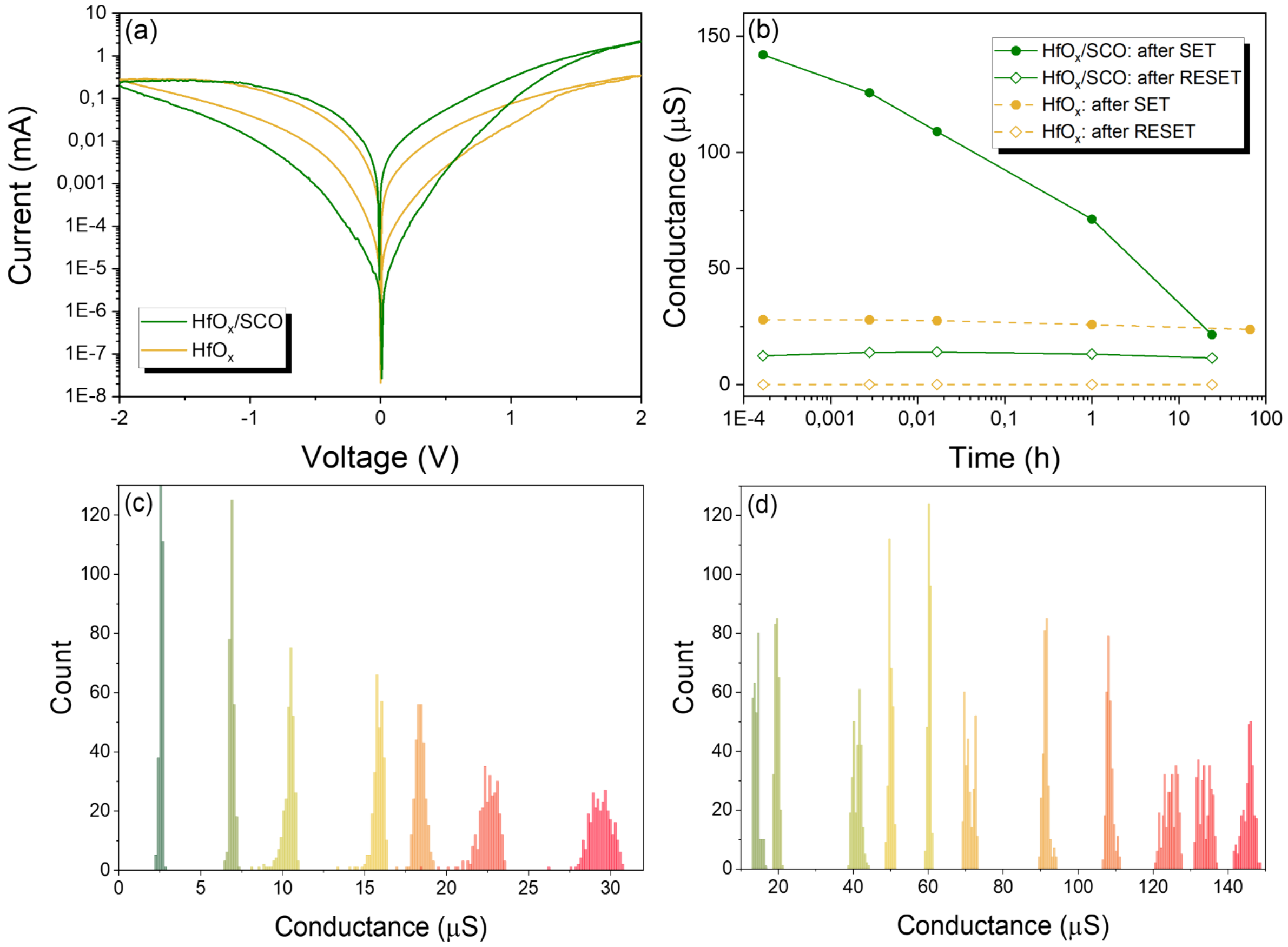

**Figure 5.** (a) Electrical characterization ($I$–$V$ curves) of the $HfO_x$ and $HfO_x$/SCO samples (220 × 220 μm$^2$ pads). (b) Measured conductance at the read voltage in the LRS (open diamonds) and HRS (filled circles) as a function of time for $HfO_x$ /SCO (dashed green line) and $HfO_x$ (solid orange line). Current histograms showing distinguishable resistance states for the (c) $HfO_x$ and (d) $HfO_x$/SCO samples.

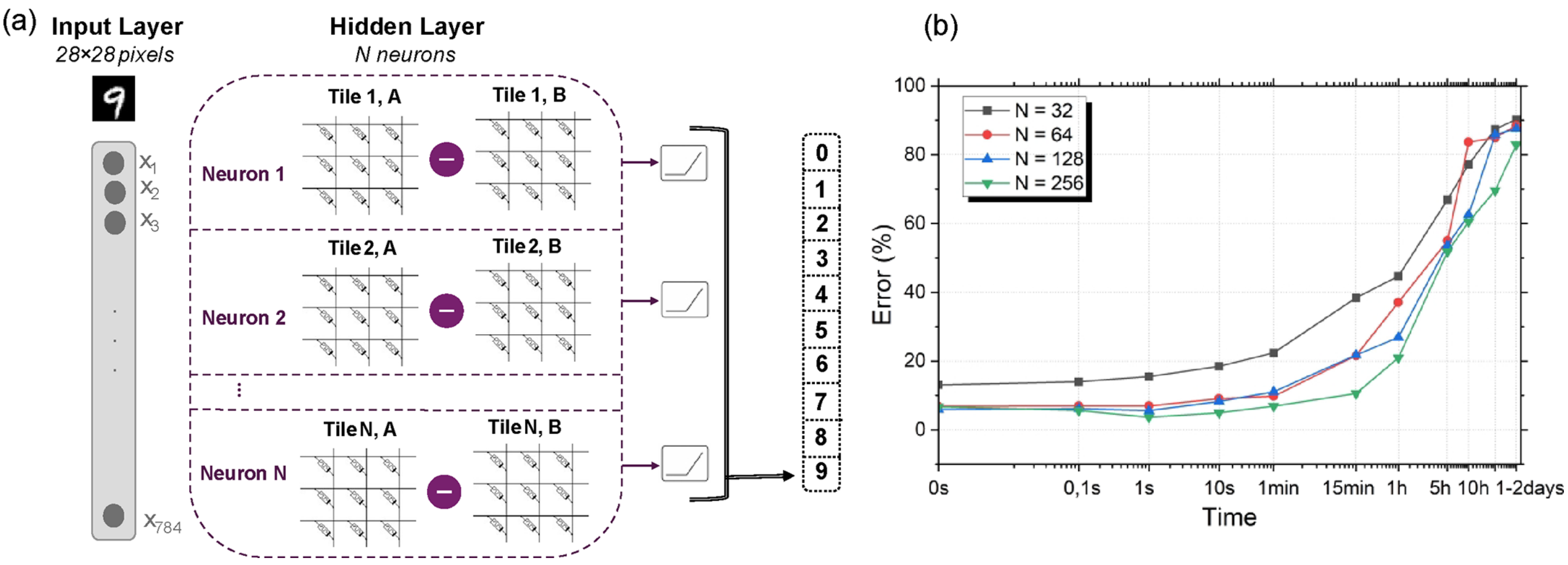

**Figure 6.** (a) Schematic representation of the used neural network, where $N \in \{32, 64, 128, 256\}$ is the hidden layer size. (b) Classification error on the MNIST dataset as a function of time after programming for hidden-layer sizes. The same baseline network was programmed once and then allowed to drift to each time point without reprogramming.

As shown in Figure 5(b), we monitored the relaxation of the system from the HRS and LRS, both for $HfO_x$ and $HfO_x$/SCO

devices. Similar to STO/SCO, a pronounced drift toward the HRS was observed for the $HfO_x$/SCO device, accompanied by an almost fivefold increase in the width of the memory window. The $HfO_x$ device, on the other hand, showed more stable states but proved significantly more fragile, often failing after routine characterization cycles. This behavior further supports the notion that the improved interface in the $HfO_x$/SCO heterostructure enhances memristor endurance.

Furthermore, similarly to the STO case, the state distribution histograms reveal a significant increase in the number of distinguishable resistance states [Figure 5(c) and (d) for $HfO_x$ and $HfO_x$/SCO, respectively]. However, the switching behavior differs markedly between the STO- and $HfO_x$-based devices. In the STO devices, the analog switching behavior enables the use of a write-verify methodology to program virtually any conductance state. In contrast, the $HfO_x$ devices exhibit discrete, metastable conductance states. Even when an intermediate conductance value is intentionally programmed, the device rapidly relaxes to the nearest stable state. We attribute this behavior to the amorphous nature of the $HfO_x$ films, in which switching is likely governed by the formation of a discrete number of conductive filaments, giving rise to metastable conductance levels. This comparison demonstrates the transferability of the proposed interface-engineering strategy across different memristive material systems, even under distinct filament formation mechanisms.

### E. Performance of STO/SCO Memristor Crossbar Array for MNIST Classification Tasks

The observed time evolution of the resistance states can be leveraged in applications such as spiking neural networks or reservoir computing. In fact, even in conventional transistor-based in-memory computing architectures, multiplier arrays are commonly reprogrammed,[27] making memristors very long retention times a requirement mostly for data storage, with the added advantage of supporting multiple quantized states beyond conventional binary transistor technology.[28] In this work, we focus on the use of memristors as multipliers, aiming to understand the trade-offs between achieving many distinguishable resistance states and maintaining their temporal stability.

To evaluate the performance of the fabricated STO/SCO memristors, we extracted the time evolution of their conductance states and the corresponding standard deviations from a statistical set of ten devices, as reported elsewhere.[29] The networks were mapped to analog form using IBM's AIHWKit,[30] which simulates the experimental behavior of the STO/SCO-based memristive devices. The measured drift and variability were incorporated through a custom ReRAM noise model.[31]

The network structure consisted of 784 input pixels ($28 \times 28$ images) fully connected to $N$ hidden neurons, followed by a fully connected output layer with 10 neurons. Here, each neuron corresponds to a weight implementing a ReLU function. In our implementation, each weight was mapped to a differential conductance pair ($G+$,$G-$), corresponding to two memristors. This differential scheme reduces the impact of common-mode conductance shifts, device-to-device asymmetries, nonlinear conductance–voltage behavior, and readout offsets.[32] Consequently, the physical realization of the network requires either the fabrication of 2N tiles [see the schematic representation in Figure 6(a)] or, alternatively, the reprogramming of the memristor arrays, analogous to the approach commonly adopted in conventional electronic systems.

Five conductance levels between 12 and 60 μS mapped linearly to the digital weights. The use of five conductance states was chosen as a compromise between the experimentally available number of distinguishable states and their temporal stability. Increasing the number of conductance levels reduces the initial error rate due to the advantages of multibit operation, but leads to faster overlap between states and, consequently, a sharper decay of the operation window. Conversely, reducing the number of states improves temporal robustness at the expense of a higher initial error rate (see Figure S7, Supporting Information). The five-state implementation was therefore selected as a reasonable compromise between computational precision and temporal stability.

The hidden layer size was varied from $N = 32$ to $N = 256$ neurons to explore the trade-off between model capacity and the network size. We then trained the neural network, parametrized by the experimental data, using the MNIST handwritten-digit dataset.[33] Training was performed digitally in PyTorch using quantized weights limited to five discrete integer levels. The quantization step was adapted to the standard deviation of each layer's weights to promote an effective occupation of the available levels. The training dataset (60,000 images) was augmented using random affine transformations, including rotations up to ±15°, translations and scaling up to 10%, and shear up to ±10°, with occasional perspective distortions to enhance generalization. The 10,000 test images were used as a fixed validation set. To improve robustness to device variability, we performed hardware-aware training on the weights by injecting Gaussian noise into the weights, with a standard deviation of 10% and an application probability of 50% per training step.

For each network, the synaptic weights were programmed once at $t = 0$ s, and their conductance drift was then simulated over time without reprogramming. Performance evaluation was carried out different timeframes, ranging from 0.1 s to 2 days. Figure 6(b) depicts the classification error as a function of time for different hidden-layer sizes. Increasing the hidden layer size reduces the initial classification error, which decreases from approximately 13% for $N = 32$ to below 7% for larger networks. These values compare favorably with state-of-the-art ferroelectric tunnel-junction (FTJ) devices, for which we also calculated an error of around 11% for the same layer size, using the reported device parameters.[34] Although the initial error is similar for hidden-layer sizes $N \geq 64$, the network with $N = 256$ maintains a lower baseline error at early times (below 1 h) compared with smaller networks. As expected, ionic drift induces progressive conductance variation and a corresponding increase in classification error. After two days the error exceeds 80% for all hidden-layer sizes, effectively rendering the system unsuitable for inference without compensation or reprogramming.

Therefore, the enlarged conductance-state window of the STO/SCO memristors can be effectively exploited for neuromorphic inference, in applications where tile reprogramming is performed within a timeframe of less than approximately ≈1 h.

## IV. CONCLUSIONS

In conclusion, we have demonstrated that introducing a high-ionic-mobility layer at the electrode/dielectric interface significantly enhances memristive device performance. Specifically, incorporating an SCO interfacial layer in STO-based memristors expanded the memory window from 8 to 22 states, reduced SET/RESET voltages by 50%, and improved endurance, albeit with reduced state retention. A two-layer fully connected neural network, trained using the experimental device characteristics, achieved classification errors below 7% on the MNIST dataset. Reliable operation requires state updates within 1 h due to temporal drift. Finally, we show that this interface-engineering strategy is broadly transferable, achieving comparable performance improvements in $HfO_x$-based memristive devices.

## ■ ASSOCIATED CONTENT

### Supporting Information

The Supporting Information is available free of charge at https://pubs.acs.org/doi/10.1021/acsaelm.6c01143.

> Comparison between the forming curves for the STO/SCO and STO devices; *I–V* curves for different contact sizes: 71, 55, and 32 μm of lateral size for the STO/SCO and STO samples; Device-to-device and cycle-to-cycle variability for the STO/SCO and STO samples; Example of current as a function of time for RESET and SET pulses, followed by a train of read pulses (50 ms, 0.5 V); SET voltage calculation: absolute current variation between the fully RESET state and after SET pulses with voltages up to 2 V and pulse widths ranging from 5 μs to 50 ms; Representative examples of multi-level memristors reported in the literature; Classification error on the MNIST dataset as a function of time after programming for different number of conductance levels (DOCX)

## ■ AUTHOR INFORMATION

### Corresponding Author

**José Diogo Costa** – *CiQUS, Centro Singular de Investigacion en Quimica Bioloxica e Materiais Moleculares, Santiago de Compostela 15782, Spain; Departamento de Quimica-Fisica, Universidade de Santiago de Compostela, Santiago de Compostela 15782, Spain;* orcid.org/0000-0001-5092-4824; Email: josediogo.teixeira@usc.es

### Authors

**Noa Varela-Domínguez** – *CiQUS, Centro Singular de Investigacion en Quimica Bioloxica e Materiais Moleculares, Santiago de Compostela 15782, Spain; Departamento de Quimica-Fisica, Universidade de Santiago de Compostela, Santiago de Compostela 15782, Spain;* orcid.org/0000-0001-9324-5936

**Daniel Veira-Canle** – *Centro Singular de Investigación en Tecnoloxías Intelixentes (CiTIUS), Universidade de Santiago de Compostela, Santiago de Compostela 15782, Spain*

**Nicholas Davey-García** – *CIC nanoGUNE BRTA, Donostia–San Sebastian 20018, Spain; Departamento de Polímeros y Materiales Avanzados: Física, Química y Tecnología, University of the Basque Country (UPV/EHU), Donostia-San Sebastián, Basque Country 20018, Spain;* orcid.org/0009-0005-5736-6366

**Victor Leborán** – *CiQUS, Centro Singular de Investigacion en Quimica Bioloxica e Materiais Moleculares, Santiago de Compostela 15782, Spain;* orcid.org/0000-0001-9806-3234

**Rafael Ramos** – *CiQUS, Centro Singular de Investigacion en Quimica Bioloxica e Materiais Moleculares, Santiago de Compostela 15782, Spain; Departamento de Quimica-Fisica, Universidade de Santiago de Compostela, Santiago de Compostela 15782, Spain;* orcid.org/0000-0001-6840-3488

**Fèlix Casanova** – *CIC nanoGUNE BRTA, Donostia–San Sebastian 20018, Spain; IKERBASQUE, Basque Foundation for Science, Bilbao 48009, Spain;* orcid.org/0000-0003-0316-2163

**Luis E. Hueso** – *CIC nanoGUNE BRTA, Donostia–San Sebastian 20018, Spain; IKERBASQUE, Basque Foundation for Science, Bilbao 48009, Spain;* orcid.org/0000-0002-7918-8047

**Victor M. Brea** – *Centro Singular de Investigación en Tecnoloxías Intelixentes (CiTIUS), Universidade de Santiago de Compostela, Santiago de Compostela 15782, Spain*

**Paula López** – *Centro Singular de Investigación en Tecnoloxías Intelixentes (CiTIUS), Universidade de Santiago de Compostela, Santiago de Compostela 15782, Spain*

**Francisco Rivadulla** – *CiQUS, Centro Singular de Investigacion en Quimica Bioloxica e Materiais Moleculares, Santiago de Compostela 15782, Spain; Departamento de Quimica-Fisica, Universidade de Santiago de Compostela, Santiago de Compostela 15782, Spain;* orcid.org/0000-0003-3099-0159

Complete contact information is available at: https://pubs.acs.org/doi/10.1021/acsaelm.6c01143

### Notes

The authors declare no competing financial interest.

## ■ ACKNOWLEDGMENTS

This work has received financial support from “Cátedra Televés en Diseño Microelectrónico” (TSI-069100-2023-0010) by the PERTE Chip, Secretaría de Estado de Telecomunicaciones e Infraestructuras Digitales, Ministerio de Asuntos Económicos y Transformación Digital and has been co-funded by the European Union-NextGenerationEU. Also, financial support from the Spanish Ministry of Science and Innovation project PID2022-138883NB-I00. Funding for open access charge: Universidade de Santiago de Compostela/CISUG. J.D.C. and R.R. acknowledge financial support from the Spanish Ministry of Science and Innovation through the Ramón y Cajal program (grants No. RYC2024-050025-I and RYC2019-026915-I, respectively). R.R also acknowledges CNS2024-154679 funded by the MCIN/AEI/10.13039/501100011033 and by the ESF investing in your future and the European Union NextGeneration EU/PRTR, the Xunta de Galicia (ED431F 2022/04, ED431C 2024/05, Centro Singular de Investigación do Sistema universitario de Galicia Accreditation 2023-2027, ED431G 2023/03) and the European Union (European Regional Development Fund - ERDF). V.B. and P.L. acknowledge funding from PID2024-155219OB-C32. N. D.-G., F.C and L.E.H. acknowledge funding from MICIU/AEI/10.13039/501100011033 (Grant No. CEX2020-001038-M) and from MICIU/AEI and ERDF/EU (Project No. PID2024-155708OBI00). N.D.-G. has been funded by the Department of Education of the Basque Government under the Predoctoral Programme for the training of nondoctoral research staff.